# Patterning of Sodium Ions and the Control of Electrons in Sodium Cobaltate

M.Roger[1], D.J.P. Morris[2], D.A. Tennant[3], M.J.Gutmann[4], J.P. Goff[2], J.-U. Hoffmann[3], R. Feyerherm[3], E. Dudzik[3], D. Prabhakaran[5], A.T. Boothroyd[5], N. Shannon[6], B. Lake[3] and P.P. Deen[7]

[1] Service de Physique de l'Etat Condense, CEA Saclay, F-91191 Gif Sur Yvette, France.
[2] Department of Physics, University of Liverpool, Oliver Lodge Laboratory, Liverpool L69 7ZE, U.K.
[3] Hahn-Meitner Institut, Glienicker Str. 100, Berlin D-14109, Germany, and Institut für Festkörperphysik, Technische Universität Berlin, Hardenbergstr. 36, 10623 Berlin, Germany.
[4] ISIS Facility, Rutherford Appleton Laboratory, Chilton, Didcot, Oxon OX11 0QX, U.K.
[5] Clarendon Laboratory, Parks Road, Oxford OX1 3PU, U.K.
[6] H.H. Wills Physics Laboratory, University of Bristol, Bristol BS8 1TL, U.K.
[7] European Synchrotron Radiation Facility, B.P 220, 38043 Grenoble Cedex, France.

**$Na_xCoO_2$ has emerged as a material of exceptional scientific interest due to the potential for thermoelectric applications[1,2], and because the strong interplay between the magnetic and superconducting properties has led to close comparisons with the physics of the high-$T_c$ cuprates[3]. The density, $x$, of the sodium in the intercalation layers can be altered electrochemically, directly changing the number of conduction electrons on triangular Co layers[4]. Recent electron diffraction measurements reveal a kaleidoscope of $Na^+$ ion patterns as a function of concentration[5]. Here we use single-crystal neutron diffraction supported by numerical simulations to determine the long-range three-dimensional superstructures of these ions. We show that the sodium ordering and its associated distortion field are governed by pure electrostatics, and the organizational principle is the stabilization of charge droplets that order long range at some simple fractional fillings. Our results provide a good starting point to understand the electronic properties in terms of a Hubbard Hamiltonian[6] taking into account the electrostatic potential from the Na superstructures. The resulting depth of potential wells in the Co layer is greater than the single-particle hopping kinetic energy. As a consequence, holes occupy preferentially the lowest potential regions and, therefore, the $Na^+$ ion patterning plays a decisive role in the transport and magnetic properties.**

Within the metallic $CoO_2$ units, the state of the conduction electrons is unconventional with large effective mass and small Fermi temperature, a high Seebeck coefficient[1], and superconductivity below 5K when hydrated[3]. Furthermore, magnetic fields strongly influence the thermopower and it has been proposed that large spin entropy, associated with the conduction electrons, causes cooling when an electric current flows under applied voltage[7]. Although the exact origin of these exotic magnetic and transport properties remains contentious, both the spin and electronic degrees of freedom of Co ions and the layered hexagonal geometry of $CoO_2$ are of key importance. Another remarkable feature of $Na_xCoO_2$ is that $Na^+$ ions sandwiched between $CoO_2$ layers are able to tunnel between different intercalation sites. The $Na^+$ layers are chargeable electrochemically and the plateaus and inflections measured in capacity-voltage[8] indicate complex behaviour. At present it is not clear whether the mechanism for sodium ordering relies mainly on Coulomb interactions[9], or more complex cooperative phenomena involving charge or orbital ordering in the Co layers[10].

Sodium ions can occupy sites between oxygen atoms in the $CoO_2$ layers, as illustrated in Fig. 1(a). These form two interpenetrating triangular lattices, denoted Na1 and Na2. Cobalt ions lie above and below Na1 sites, resulting in an extra energy cost for occupation relative to Na2 of $\Delta$, which depends purely on the effective dielectric constant $\varepsilon$. $Na^+$ ions are bigger than the distance between Na1 and Na2 sites, disallowing simultaneous occupancy of nearest neighboring sites. In addition there is electrostatic repulsion between the $Na^+$ ions consisting of a long-range Coulomb potential $e/(4\pi\varepsilon\varepsilon_0 r)$ and a short-range Na-Na ion shell repulsion V=0.06eV between neighbours on the same sublattice.

Exact Ewald summations of long-range Coulomb energies reveal that the spontaneous formation of multi-vacancy clusters drives the organization of $Na^+$ ions, as illustrated in Fig. 1(b). $Na^+$ ion vacancies become attractive at very short length scales, through a delicate balance of Coulomb forces and onsite Na1 energy, and condense into droplets. As droplet size increases, the extra energy cost of Na1 sites in the droplet core eventually cancels the energy gain of the droplet surface. Below a critical temperature $T_c$ and for simple fractional values of $x$, droplets of identical size condense in regular arrays keeping them as far apart as possible to minimize Coulomb repulsion. Above $T_c$, entropy is gained by the coexistence of different sized droplets, keeping only short-range correlations. This "glassy" droplet structure is destroyed at temperatures which exceed the stability domain of $Na_xCoO_2$.

At low concentration we find the stable phases at $x$=1/3 comprising only Na2 sites, and $x$=1/2 with an equal occupation of Na1 and Na2 sites, as observed experimentally[10,11]. This latter striped phase can also be viewed as closely packed di-vacancy clusters. Figure 1(c) plots the ground-state energy for increasing simple fractional concentrations $x$ of regular arrays of ordered mono-, di-, tri- and quadri-vacancy clusters. While di-vacancy phases are the most favorable for $0.5 < x < 0.71$, tri-vacancy phases gain more energy at $0.75 < x < 0.8$ and quadri-vacancy phases, although very close in energy to tri-vacancy phases, are marginally stable for $0.8 < x < 0.85$. There are a number of phases in the vicinity of $x \sim 0.8$ with comparable energy, and this explains the observation of a wide variety of superstructures in this composition range. In addition for $x$>0.8 coexistence of $x$=1 phase with a phase of lower $x \approx 0.8$ is predicted, as observed through NMR[12].

We performed neutron diffraction measurements on single-crystal $Na_xCoO_2$ of composition $x$=0.75, 0.78 and 0.92, see Methods. Figures 2(a-c) show the data for $x$=0.75 at $T$=150K for the hexagonal $L$=11, 10 and 9 planes, respectively, and this pattern is observed down to the lowest temperatures studied, $T$=1.5K. Rings of superlattice peaks form around the main Bragg peaks together with higher order harmonics, and these arise from long-range ordering of Na ions. There is a complex modulation of superlattice peak intensity along $L$, and within the plane around the rings of superstructure peaks.

The observed commensurate modulation wave vectors, correspond to the supercell of the $x$=0.80 tri-vacancy model, ***a'*** = ***a*** + 3***b*** and ***b'*** = 4***a*** - 3***b***, (see Supplementary Information Fig. 1) which is shown to be within a few K of the energetically most stable phase in Fig. 1(c). The Fourier transforms from this phase show good agreement with experiment, see Figs. 2(d-f). The variation of intensity within the plane rules out other mono- and di-vacancy cluster models, see Supplementary Information Fig. 2. The electrostatic potential energy is lowest when the tri-vacancy clusters in successive sodium layers are shifted by the maximum amount within the plane to the body centre of the unit cell, and the calculated scattering agrees best with the data for this model. To obtain quantitative agreement with the observed intensities further displacement of $CoO_6$ octahedra is required. The gradient of the electrostatic potential from $Na^+$ ions has been calculated at each of the Co sites and there is a force component along the hexagonal $c$-direction away from the tri-vacancy clusters. The displacement of the Co ions is assumed to be proportional to this force, and the oxygen ions move in such a way to keep the Co-O bond length constant, see Fig. 3. The model has just one adjustable parameter, the force constant, and best agreement with the data is obtained with a maximum displacement of the Co/O ions of only a 0.010(1)*$c$.

We observe a first order phase transition to the $x$=0.84 quadri-vacancy phase from Fig. 1(c) at $T$=285K, and this coincides with an anomaly in the electrical resistance, see Supplementary Information Figs. 3, 4 and 5. The superstructures have also been determined for $x$=0.78 and 0.92, and these can be understood in terms of the coexistence of phases at low temperature and the quadri-vacancy phase at high temperature. The fractional occupation of Na sites calculated in our model agrees with the available neutron diffraction data[13,14,15] over a wide range of $x$, see Supplementary Information Fig. 6.

We clearly identify two energy scales: the electrostatic energy driving the formation of vacancy clusters (~500meV), and a much lower energy from the deformation of the $CoO_2$ layer (a few meV). The good agreement of the experimental results with the vacancy clustering predicted by our first-order treatment opens new perspectives on the study of elastic deformations and cooperative interactions with mobile carriers in the $CoO_2$ layers using perturbative methods. The approach may be applied to other transition-metal oxides with ions partially filling spaces in hexagonal planes.

The $Na^+$ ion patterning determines the Coulomb landscape in the Co planes, and this depends very sensitively on the Na superstructure. The electronic properties may be modeled in terms of a modified Hubbard model

$$H = -t \sum_{\langle i,j \rangle \sigma =\uparrow,\downarrow} c_{i\sigma}^+ c_{j\sigma} + U \sum_i n_{i\uparrow} n_{i\downarrow} + \sum_i E_i \left( n_{i\uparrow} + n_{i\downarrow} \right)$$

where $U$ is the penalty for double occupation of a Co site, and the $E_i$'s are on-site energies corresponding to different positions of the Co's with respect to the modulation of the Coulomb landscape in the supercell, induced by the ordered array of Na ions.

Figure 4 shows two contrasting cases, $x$=0.5 and 0.80. Coulomb-potential stripes are obtained for $x$=0.5 in agreement with the low and high-spin charge-ordered stripes found using neutron diffraction[16]. Potential wells of depth ~100meV are predicted for $x$=0.80. Since the total bandwidth of $Na_{0.7}CoO_2$ is less than 100 meV and the single-particle hopping frequency is about $t\sim10$ meV[17], the electrostatic potential may be expected to localize holes. This picture is consistent with the multiple valence states detected by NMR[18], and the conduction paths around the localized regions readily account for the observed metallic behaviour. Furthermore, the fact that the separation between localized regions of spin within the plane is comparable to the spacing between planes explains the 3D magnetism revealed by the magnetic excitations[19,20].

Notably multi-vacancy clusters form cages in which Na ions can vibrate relatively freely. These would be expected to disrupt the propagation of phonon excitations, leading to a low thermal conductivity. The "rattling" of cations inside lattice cavities has been observed in a range of promising candidates for thermoelectric applications[21,22,23]. This behaviour is confirmed for metallic $Na_xCoO_2$ by the reported low temperature-independent thermal conductivity[11]. The rare coincidence of high electrical conductivity and low thermal conductivity are precisely the conditions required for high figure of merit thermoelectric materials.

Finally, while the mechanism of superconductivity in the hydrated phases will also depend upon other factors, sodium clustering will have an important effect. The electrostatic influence of Na ordering on the $CoO_2$ layers will be extremely sensitive to the intercalation of dipolar water molecules. In bi-hydrate $Na_xCoO_2 \cdot yH_2O$ water molecules intercalate *between* the $Na^+$ ions and the $CoO_2$ planes, and therefore screen the conduction electrons from the Na potential. Superconductivity is observed at 5K[3] consistent with effective screening of the pair-breaking Coulomb potential. In contrast for mono-hydrates the water molecules are intercalated *within* the planes of $Na^+$ ions and cannot screen the $CoO_2$ layers effectively, and indeed, like their anhydrous counterparts, these materials fail to superconduct[24].

In summary, elaborate sodium ordering in $Na_xCoO_2$ has been measured using neutron scattering. Organization of the sodiums proceeds by formation of vacancy clusters which order in a periodic lattice. A model to explain the results is proposed based on ionic shell and Coulomb interactions between sodiums. The ordered sodium lattice influences the metallic properties of the $CoO_2$ layers via its periodic Coulomb potential and exerts control over many of the thermoelectric, magnetic and superconducting properties. Our results show that $Na_xCoO_2$ is a model material in which to investigate electrochemical control of magnetic and electronic properties.

**Methods**

**Experiment**

Measurements were made from samples obtained from zone-melted rods of $Na_{0.75}CoO_2$, $Na_{0.78}CoO_2$ and $Na_{0.92}CoO_2$ grown using the floating zone method[25]. Single crystals from the rods were initially screened and a suitable piece in each case cut away.

Neutron Laue diffraction provides a powerful technique with which to probe atomic ordering in the bulk of a crystal sample. Single-crystal data were collected on the SXD diffractometer at the ISIS pulsed neutron source, Rutherford Appleton Laboratory, UK. SXD combines the white beam Laue technique with area detectors covering a solid-angle of $2\pi$ steradians allowing comprehensive data sets to be collected. Samples were mounted on aluminium pins and cooled to 150 K using a closed-cycle He refrigerator. A typical data set required four to five orientations to be collected for 12 to 35 hours per orientation depending on sample size. Data were corrected for incident flux using a null-scattering V/Nb sphere. These data were then combined to a volume in reciprocal space and sliced to obtain individual planar cuts.

Additional measurements were performed using the flat-cone diffractometer E2 at the Hahn-Meitner Institute in Germany. E2 was used to determine accurate modulation wave vectors. Measurements were performed using a variable-temperature cryostat down to a base temperature of 1.5 K and the temperature dependence was determined through the phase transition. Other small samples from the same rods were examined by synchrotron x-ray diffraction using the MAGS beamline at BESSY in Berlin, and electrical transport and magnetometry measurements were performed to assess their quality, see Supplementary Information Figs. 3 and 5.

The structural superlattice neutron diffraction peaks are typically two orders of magnitude lower in intensity than the hexagonal Bragg peaks. The difficulty in observing the main magnetic Bragg reflections suggests that detection of any in-plane modulation of the magnetic structure would be difficult, even with polarized neutrons.

The quoted sodium concentrations, $x$, were determined using EPMA. We note that neutron activation analysis gave a value of $x$ roughly 5% lower. Small inclusions of a few percent of the cobalt oxides (CoO and $Co_3O_4$) as well as $Na_2O$ were found. The diffraction measurements show that these grow epitaxially on the host lattice and the impurity signal is straightforward to distinguish from that from the $Na_xCoO_2$.

**Theory**

The pair energy between $Na^+$ ions is the sum of a strongly repulsive short-range part preventing electronic shells from overlapping $V_{sr} = A\exp[-r/r_0] - C/r^6$ with A=424 eV, $r_0$= 0.318 Å and C= 1.05 eV Å$^6$ from the literature[26] and a long-range Coulomb contribution $V_c = e^2/4\pi\varepsilon\varepsilon_0 r$. At nearest and next-nearest neighbor distances, we obtain respectively $V_{sr}^1 = 2.53\,eV$ (considered as infinite at room temperature) and $V_{sr}^2 = 0.06\,eV$ (we neglect $V_{sr}$ at larger distances). The Coulomb landscape in Fig. 4 depends on one parameter: the effective dielectric constant $\varepsilon$. We take it to be isotropic and use the value $\varepsilon = 6$ to ensure stability of the long-range ordered phases up to room temperature. Note that our conclusions concerning vacancy clustering and the importance of the potential-well depth compared to hopping energy are quite robust and remain valid within a very large $\varepsilon$ range.

Long-range Coulomb interactions between all Co, O and Na ions on three-dimensional periodic lattices were summed with an accuracy of 10 digits using a generalization of the Ewald method[27]. At finite temperature, Monte-Carlo simulations were performed in Canonical and Grand-Canonical ensembles on periodic lattices including up to 1800 Na sites. Vacancy clustering has been observed in the whole stability range 0 < T < 600 K of $Na_xCoO_2$. For a variety of simple fractional fillings, long-range order of identical clusters appears a low temperature, and this melts into a glassy mixture of di-, tri- and quadri-vacancy clusters above $T\sim400K$. Full details will be reported elsewhere.

**References**

1. Terasaki, I., Sasago, Y. & Uchinokura, K. Large thermoelectric power in $NaCo_2O_4$ single crystals. *Phys. Rev. B* **56**, R12685- R12687 (1997).
2. Lee, M. *et al*. Large enhancement of the thermopower in $Na_xCoO_2$ at high Na doping. *Nature. Mater.* **5**, 537-540 (2006).
3. Takada, K., *et al*. Superconductivity in two-dimensional $CoO_2$ layers. *Nature* **422**, 53-55 (2003).
4. Delmas, C., *et al*. Electrochemical intercalation of sodium in $Na_xCoO_2$ bronzes. *Solid State Ionics* **3-4** 165-169 (1981).
5. Zandbergen, H. W. *et al*. Sodium ion ordering in $Na_xCoO_2$: Electron diffraction study. *Phys. Rev. B* **70** 024101 (2004).
6. Hubbard, J., Electron correlations in narrow energy bands. *Proc. R. Soc. London A* **276**, 238 (1963).
7. Wang, Y.Y., Rogado, N.S., Cava, R.J. & Ong, N.P. Spin entropy as the likely source of enhanced thermopower in $NaxCo_2O_4$. *Nature* **423**, 425-428 (2003).
8. Amatucci, G. G., Tarascon, J. M. & Klein, L. C. $CoO_2$, The End Member of the $Li_xCoO_2$ Solid Solution. *J. Electrochem. Soc.* **143**, 1114-1123 (1996).
9. Zhang, P., Capaz, R. B., Cohen, M. L. & Louie, S. G. Theory of sodium ordering in $Na_xCoO_2$. *Phys. Rev. B* **71**, 153102 (2005).
10. Huang, Q., *et al*. Low temperature phase transition and crystal structure of $Na_{0.5}CoO_2$. *J. of Phys. Cond. Mat.* **16,** 5803-14 (2004).
11. Foo, M. L., *et al*. Charge Ordering, Commensurability, and Metallicity in the Phase Diagram of the Layered $Na_xCoO_2$. *Phys. Rev. Lett*. **92**, 247001 (2004).
12. de Vaulx, C. *et al*. Nonmagnetic insulator state in $Na_1CoO_2$ and phase separation of Na vacancies. *Phys. Rev. Lett.* **95**, 186405 (2005).
13. Huang, Q., *et al*. Coupling between electronic and structural degrees of freedom in the triangular lattice conductor $Na_xCoO_2$. *Phys. Rev. B* **70**, 184110 (2004).
14. Balsys, R. J. and Davis, R. L. Refinement of the structure of $Na_{0.74}CoO_2$ using neutron powder diffraction. *Solid State Ionics* **93**, 279 (1996).
15. Mukhamedshin, I. R., *et al*. $^{28}$Na NMR Evidence for Charge Order and Anomalous magnetism in $Na_xCoO_2$. *Phys. Rev. Lett.* **93**, 167601 (2004).
16. Yokoi, M. *et al*, Magnetic Correlation of $Na_xCoO_2$ and Successive Phase Transitions of $Na_{0.5}CoO_2$ – NMR and Neutron Diffraction Studies. *J. Phys. Soc. Jpn.* **74**, 3046 (2005).
17. Hasan, M. Z., *et al*. Fermi surface and quasiparticle dynamics of $Na_{0.7}CoO_2$ investigated by angle-resolved photoemission spectroscopy. *Phys. Rev. Lett.* **92,** 246402 (2004).
18. Mukhamedshin, I. R., Alloul, H., Collin, G. & Blanchard, N. $^{59}$Co NMR study of the Co states in superconducting and anhydrous cobaltates. *Phys. Rev. Lett.* **94**, 247602 (2005).
19. Bayrakci, S. P., *et al*. Magnetic Ordering and Spin Waves in $Na_{0.82}CoO_2$. *Phys. Rev. Lett*. **94,** 157205 (2005).

20. Helme, L. M., *et al*. Three-Dimensional Spin Fluctuations in $Na_{0.75}CoO_2$. *Phys. Rev. Lett*. **94**, 157206 (2005).
21. Slack, G. A. and Tsoukala, V. G. Some properties of semiconducting $IrSb_3$. *J. Appl. Phys.* **76**, 1665 (1994).
22. Nolas, G. S., *et al.* Semiconducting Ge clathrates : Promising candidates for thermoelectric applications. *Appl. Phys. Lett.* **73**, 178 (1998).
23. Nolas, G.S., *et al*. Effect of partial void filling on the lattice thermal conductivity of skutterudites. *Phys. Rev. B* **58**, 164 (1998).
24. Sakurai, H., *et al*. The role of the water molecules in novel superconductor, $Na_{0.35}CoO_2 \cdot 1.3H_2O$. *Physica C: Superconductivity*, **412-414**, 182-186 (2004).
25. Prabhakaran, D., Boothroyd, A. T., Coldea, R. & Charnley, N. R. Crystal growth of $Na_xCoO_2$ under different atmospheres. *J. Crystal Growth* **271**, 74-80 (2004).
26. Fumi, F. G. & Tosi, M. P. Ionic Sizes and Born Repulsive Parameters in the NaCl-Type Alkali Halides. *J. Phys. Chem. Solids* **25**, 31-52 (1964).
27. Sperb, R. An Alternative to Ewald Sums. *Molecular Simulation* **20**, 179-200 (1998).

**Supplementary Information** is linked to the online version of the paper at www.nature.com/nature.

**Acknowledgements**
We wish to thank Steve Lee and John Irvine of St Andrews University, Paolo Radaelli and Aziz Daoud-Aladine of the Rutherford Appleton Laboratory, Klaus Kiefer and Dimitri Argyriou from HMI Berlin, Radu Coldea of the University of Bristol and Minoru Nohara of the University of Tokyo for helpful discussions, and Themis Bowcock and Andrew Washbrook for the use of the MAP2 supercomputer at the University of Liverpool.

**Author Information**
The authors declare that they have no competing financial interests. Correspondence and requests for materials should be addressed to M.R. (e-mail: roger@drecam.saclay.cea.fr).

**Figure Legends**

**Figure 1: $Na_xCoO_2$ has $Na^+$ ions intercalated between $CoO_2$ layers**. (a) Two interpenetrating hexagonal lattices of intercalation sites denoted Na1, and Na2 in a repeated Star-of-David motif[1] are formed. Ions are shown to scale. (b) Energy for two vacancies decreases with increasing distance as expected for Coulomb repulsion. Neighbouring vacancies can reduce their energy by promotion of a Na2 sodium to the central Na1 site. The resultant *di-vacancy* cluster has net charge $2e^-$ spread over three sites and substantially lower energy as central sodium is now further from its neighbours. Its stabilization energy is lower than vacancies 3-4 sites apart, and so should form spontaneously at modest concentrations. The formation of tri-vacancies and quadri-vacancies follows a similar process. (c) Ground-state energies of superstructures for mono-vacancies (black), di-vacancies (red), tri-vacancies (blue) and quadri-vacancies (green) [a function is subtracted to show energy differences on a meV scale]. The inset shows the $x=0.80$ tri-vacancy phase.

**Figure 2: Neutron diffraction showing bulk 3D ordering of $Na^+$.** (a) $(H,K,11)$, (b) $(H,K,10)$, (c) $(H,K,9)$, planes for $x=0.75$ at $T=150K$, superlattice reflections form rings around hexagonal Bragg reflections, and some higher order harmonics are observed. (d-f) The corresponding scattering calculated using the $x=0.80$ tri-vacancy phase shown in Fig. 3. The lattice is averaged over all domains. The scattering calculated using a one-parameter model (see text) captures the observed modulation both within and between planes. The cancellation of intensity due to interference between regular oxygen sites makes the scattering at $L=11$ particularly sensitive to the sodium ordering and associated distortion field.

**Figure 3: Real space superstructure.** The unit cell for the $x=0.80$ tri-vacancy superstructure used to model the low temperature phase for $x=0.75$ showing the in-plane displacement of tri-vacancies in successive Na layers. The colour scale on the Co ions shows the electrostatic potential gradient. The maximum displacement of cobalt ions in this figure and the separation between planes are increased to aid visualization. The O ions are omitted for clarity, but their positions follow the Co ion distortion since the Co-O bond length remains fixed. This leads to a small in-plane displacement of the O ions causing a small in-plane movement of Na ions [out-of-plane displacement of the Na ions is ruled out by symmetry].

---

[1] This minimal model for the sodium system can be mapped onto a frustrated Ising model where a pseudo-spin state $I_i^z = +1/2$ denotes that the $i^{th}$ site of the honeycomb lattice is occupied, and $I_i^z = -1/2$ that it is empty. The electrostatic and ionic shell interactions are represented by the Hamiltonian:

$$H = \sum_{ij}(V_{ij} + c(r_{ij}))(I_i^z + 1/2)(I_j^z + 1/2) + \sum_i(-\mu + \Delta(1+(-1)^i))(I_i^z + 1/2),$$

where only shell repulsions $V_{ij}$ for nearest neighbour and next-nearest neighbour interactions are treated as important, $c$ absorbs the Coulomb repulsions. The chemical potential $\mu$ acts as a uniform field, and Na1 onsite energy $\Delta$ like a staggered field.

**Figure 4: Coulomb potential in the Co plane calculated using ordered superstructures.** Positions of Co ions are shown as green circles, and the colour scale is in eV. Sodium superstructures are: (a) Di-vacancy array for $x=1/2$. The depth of the potential wells is of order 60 meV. Holes should localise along white stripes in Co planes. The observed striped magnetic order is shown as blue arrows[11]. (b) Tri-vacancy array for $x=0.80$, as shown in Fig. 3. $Co^{4+}$ is expected at the minima and $Co^{3+}$ at the maxima. Spin-½ holes are localized in potential wells of depth 100 meV, where localized magnetic moments are expected.

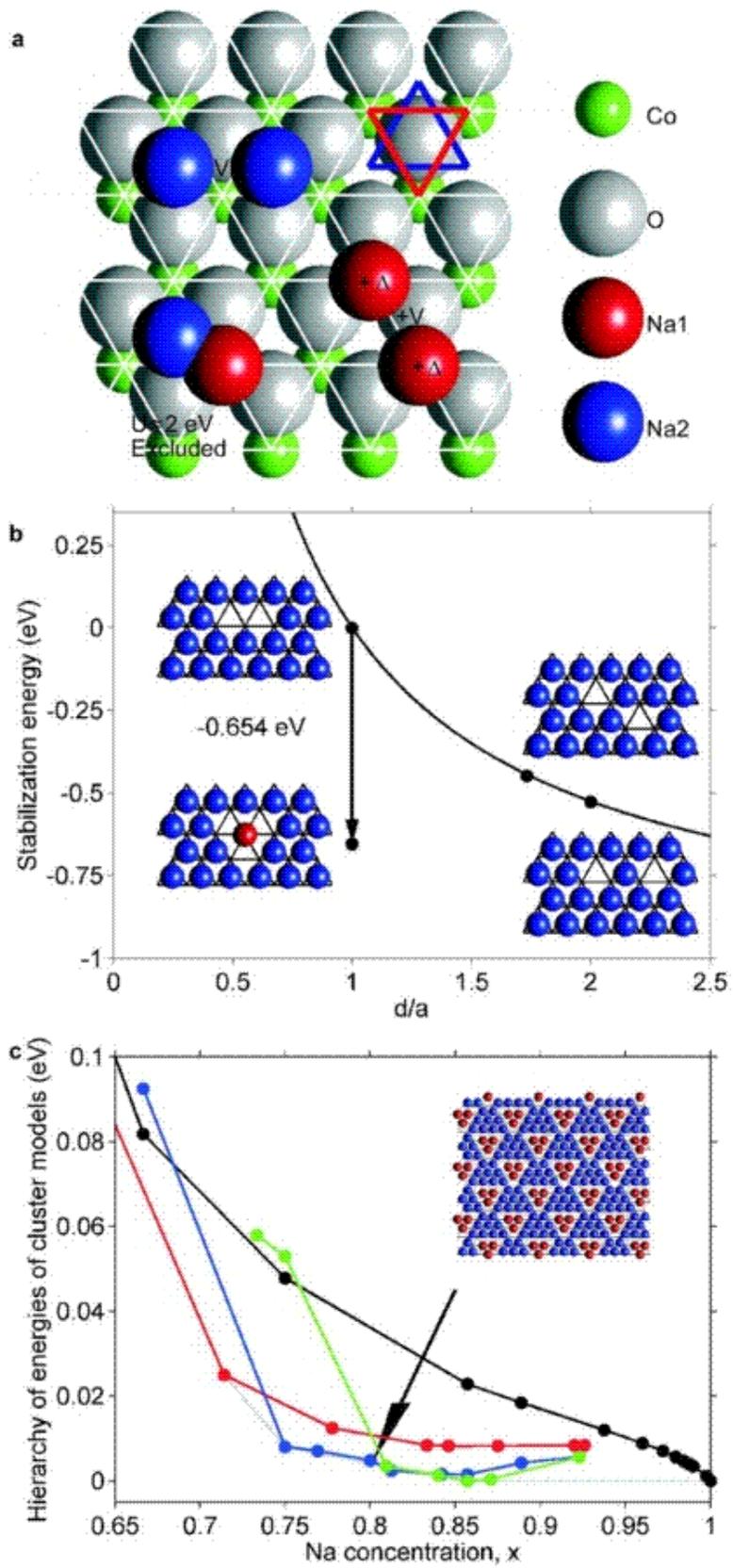

**Figure 1**

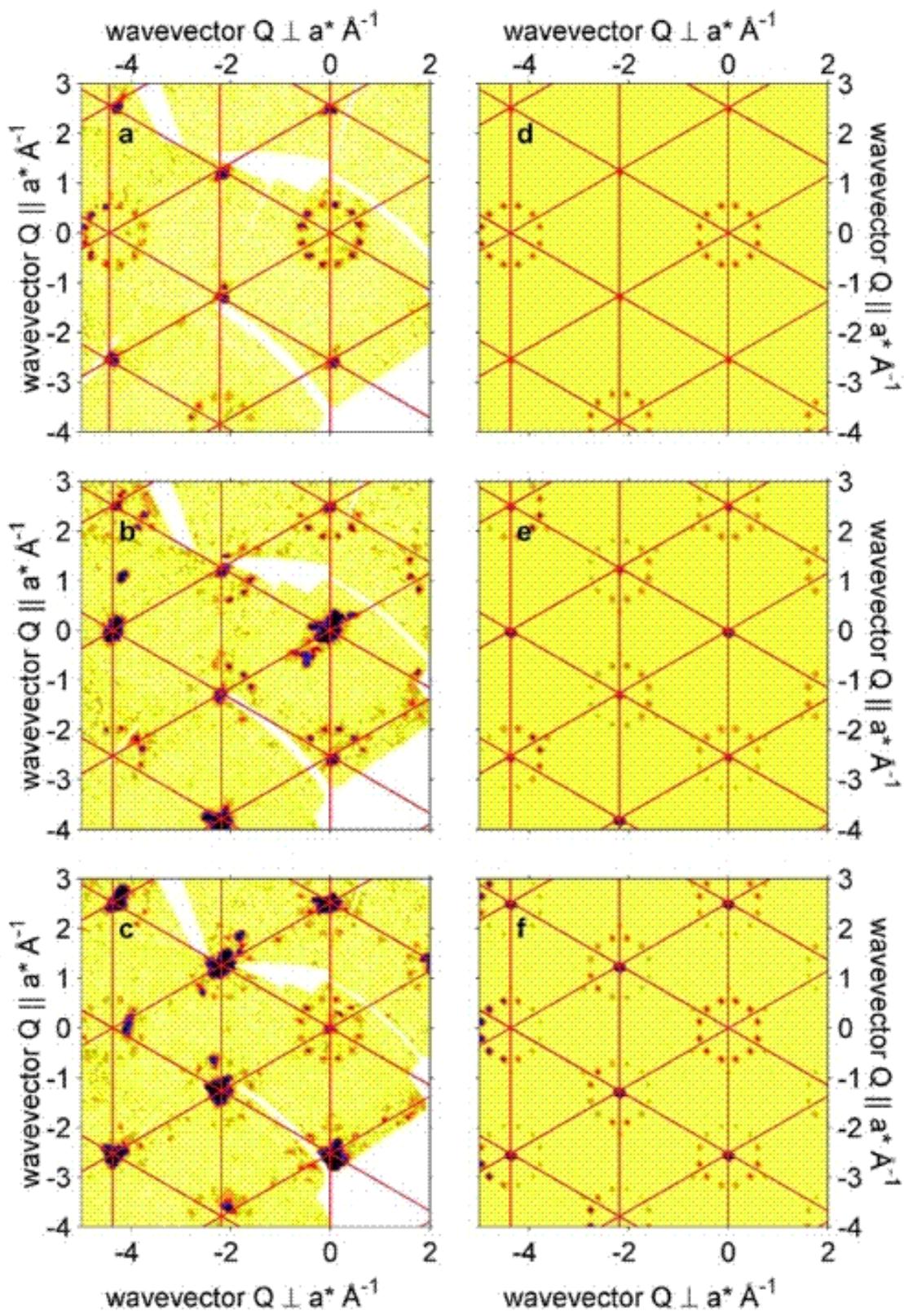

**Figure 2**

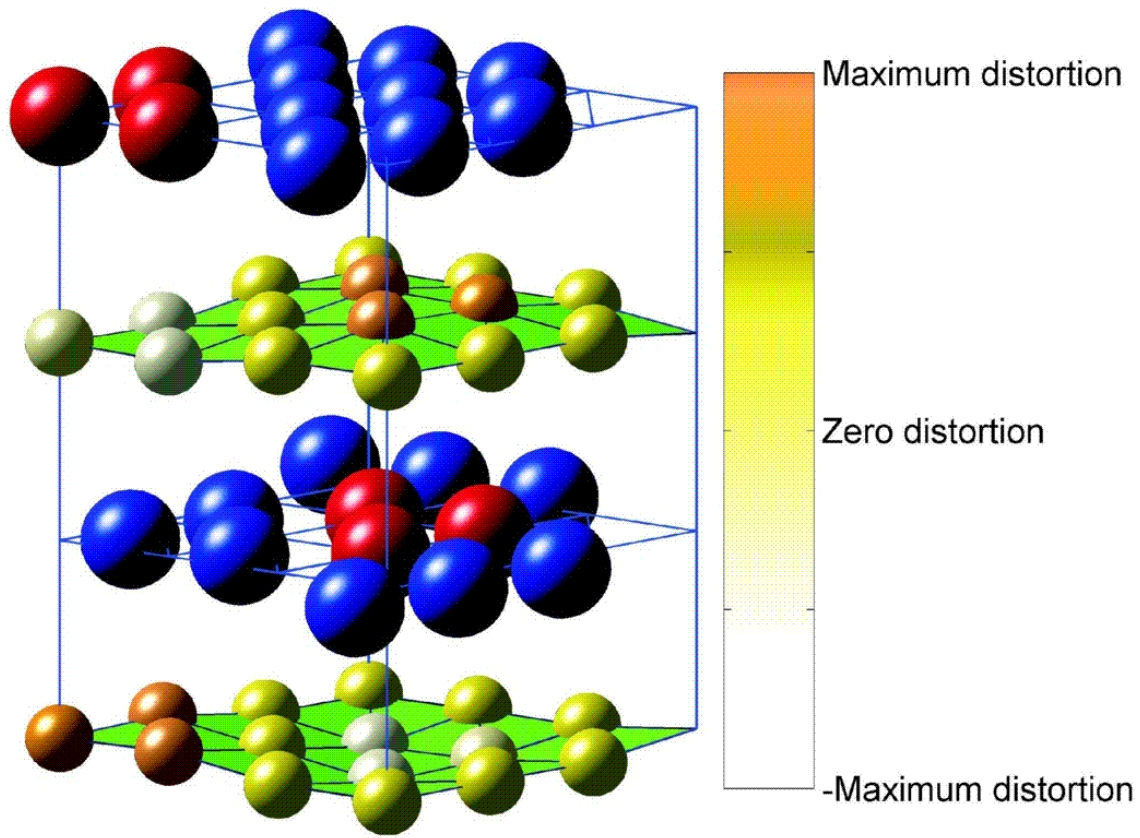

**Figure 3**

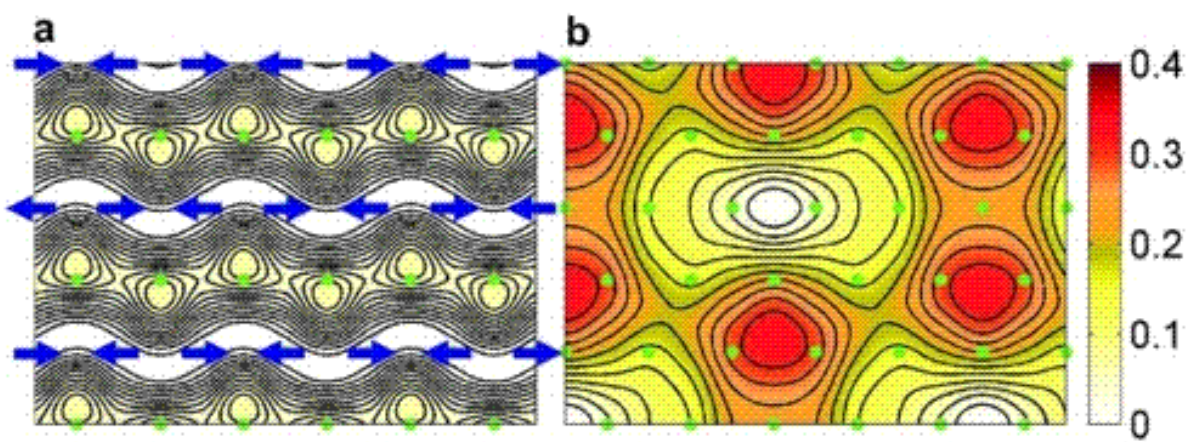

**Figure 4**